\documentclass[twocolumn]{svjour3}

\usepackage{etoolbox}
\usepackage{bbold}
\usepackage{amssymb,amsmath,amsthm}
\usepackage{mathtools} 
\usepackage{balance}
\usepackage{graphicx}
\usepackage{dcolumn}
\usepackage{bm}
\usepackage{braket}
\usepackage{color}
\usepackage{hyperref}
\usepackage{ulem,url}
\usepackage{longtable}
\usepackage{multirow}
\usepackage{makecell, cellspace}
\usepackage{array}
\newcolumntype{L}[1]{>{\raggedright\let\newline\\\arraybackslash\hspace{0pt}}m{#1}}
\newcolumntype{C}[1]{>{\centering\let\newline\\\arraybackslash\hspace{0pt}}m{#1}}
\newcolumntype{R}[1]{>{\raggedleft\let\newline\\\arraybackslash\hspace{0pt}}m{#1}}

\usepackage{balance}

\newcommand{\THETAVEC}[1]{\ensuremath{\overrightarrow{\theta}_{#1}}}
\journalname{QUMI}
\begin{document}
\title{Mode connectivity in the loss landscape of parameterized quantum circuits \thanks{This manuscript has been authored by UT-Battelle, LLC under Contract No. DE-AC05-00OR22725 with the U.S. Department of Energy. The United States Government retains and the publisher, by accepting the article for publication, acknowledges that the United States Government retains a non-exclusive, paid-up, irrevocable, worldwide license to publish or reproduce the published form of this manuscript, or allow others to do so, for United States Government purposes. The Department of Energy will provide public access to these results of federally sponsored research in accordance with the DOE Public Access Plan. (http://energy.gov/downloads/doe-public-279 access-plan).}}

\titlerunning{Mode connectivity of PQCs}
\author{Kathleen~E.~Hamilton \and
        Emily Lynn \and 
        ~Raphael~C.~Pooser}

\institute{K. E.~ Hamilton and R. C. ~Pooser \at
              Computational Science and Engineering Division \\
              Oak Ridge National Laboratory, TN, USA       \\
              \email{E-mail: hamiltonke@ornl.gov} \\
           \and
           Emily Lynn  \at
              Department of Physics and Astronomy,\\
              Taylor University, IN, USA 
}
\date{Received: date / Accepted: date}
\maketitle

\begin{abstract}
Variational training of parameterized quantum circuits (PQCs) underpins many workflows employed on near-term noisy intermediate scale quantum (NISQ) devices.  It is a hybrid quantum-classical approach that minimizes an associated cost function in order to train a parameterized ansatz.  In this paper we adapt the qualitative loss landscape characterization for neural networks introduced in \cite{goodfellow2014qualitatively,li2017visualizing} and tests for connectivity used in \cite{draxler2018essentially} to study the loss landscape features in PQC training.   We present results for PQCs trained on a simple regression task, using the bilayer circuit ansatz, which consists of alternating layers of parameterized rotation gates and entangling gates.  Multiple circuits are trained with $3$ different batch gradient optimizers: stochastic gradient descent, the quantum natural gradient, and Adam.  We identify large features in the landscape that can lead to faster convergence in training workflows.
\end{abstract}

\section{Introduction}
\label{sec:intro}
Variational training of parameterized quantum circuits (PQCs) shares a number of similarities with deep learning \cite{benedetti2019parameterized}. While a full understanding of training dynamics for deep learning models remains an open question, there have been several studies which explore the interplay of over-parameterized networks, gradient descent methods, and the number of minima in the loss landscape \cite{kawaguchi2016deep,safran2018spurious,du2018gradient,du2019gradient}. The success of gradient-based training of over-parameterized models has led to the success of deep learning in today's artificial intelligence. For variational quantum algortihms, understanding the optimization landscape associated with training workflows is necessary for scaling up quantum algorithms in the current noisy intermediate scale quantum computing (NISQ) era.  For example, since the identification of the barren plateau concept in \cite{mcclean2018barren}, a number of papers have posited that barren plateaus may exist due to cost function structure \cite{cerezo2021cost}, noise induction \cite{wang2020noiseinduced}, or as a result of excess entanglement \cite{marrero2020entanglement}. This flattening of the loss landscape is a significant obstacle to building large quantum machine learning applications which utilize gradient-based optimization.  Yet gradient based optimization remains promising as it may result in a speedup of variational training workflows \cite{napp2021gradient}. 

Previous studies of deep learning have focused on establishing whether the optimization landscape is dominated by local minima \cite{antenucci2019glassy,baity2018comparing}, the connection to spin glass models \cite{choromanska2015loss}, existence of barriers \cite{draxler2018essentially}, and understanding how network architecture affects the landscape \cite{baldassi2020shaping}. Connectivity has also been explored in the quantum control literature, with many works focusing on whether landscapes are trap-free \cite{russell2016quantum}, or if optimization can get trapped \cite{larocca2018quantum,larocca2018quantum}.  In this work we identify connected minima in the PQC landscape using the nudged elastic band algorithm (NEB) \cite{henkelman2000neb} which has been used to analyze the loss landscape of deep neural networks \cite{draxler2018essentially,garipov2018loss}.  These methods have demonstrated how low-dimensional representations of high-dimensional parameter spaces provide insight into what characteristics make training difficult or simple for certain problem configurations. Thus, knowledge of interconnectivity in the lower dimensional landscapes can inform the potential performance of a given piece of hardware, circuit, and optimizer combination, including aiding PQC design. 

We compare the performance of three commonly used gradient-based optimization methods: Quantum natural gradient \cite{stokes2020quantum}, standard stochastic gradient descent \cite{goodfellow2016deeplearning}, and Adam \cite{kingma2014adam}. We execute a high number of training runs of small, shallow circuits in noiseless simulation. The trained models and visualization methods found in the classical machine learning literature \cite{goodfellow2014qualitatively,li2017visualizing} are used to characterize the loss landscape of specific circuit designs.  

We use classical machine learning methods to characterize the PQC loss landscape. In particular we adapt visualization techniques \cite{goodfellow2014qualitatively,li2017visualizing,garipov2018loss} to generate contour plots using a hyper-plane which contains multiple minima found by different training methods. Then, we use NEB to analyze the connectivity between these minima. Analogous to studies of classical neural networks, we show that multiple minima with low MSE error can sit in connected regions of the loss landscape \cite{garipov2018loss,draxler2018essentially}. Finally, we use features of the loss landscape to identify dropout stable patterns \cite{shevchenko2020landscape} in several PQCs.

The circuits in the present study have fixed width of three, with $\leq 15$ parameters, and are trained as regression models using quantum circuit learning (QCL) \cite{mitarai2018quantum}.  The QCL training workflow is studied with noiseless qubit simulation implemented in PennyLane \cite{bergholm2020pennylane}. This means that the differences in training performance between experiments stem from optimizer choice, ansatz design, and/or hyperparameter choices (batch size and initialization). We extract sets of unique minima in the loss landscape from the trained models and demonstrate that connected paths of low loss can exist between them.  These results provide a ``quantum'' analogy to the interconnectivity present in classical deep learning models~\cite{garipov2018loss,draxler2018essentially}.  

In Section \ref{sec:theory} we provide details of our training workflows, visualizations methods, and analysis of mode connectivity. We present our QCL results in Section \ref{sec:qcl}, discuss the mode connectivity in landscapes in Section \ref{sec:connectivity}, and analyze parameter dropout stability in Section \ref{sec:dropout}. Section \ref{sec:dicussion} discusses these results and analyzes the information that we can extract from visualization of the loss landscape. In particular we focus on the prevalence of connected regions of low loss.

\section{Methods}
\label{sec:theory}

\subsection{PQC Notation}
\label{sec:notation}
\begin{figure}[htbp]
  \centering
  \includegraphics[width=\columnwidth]{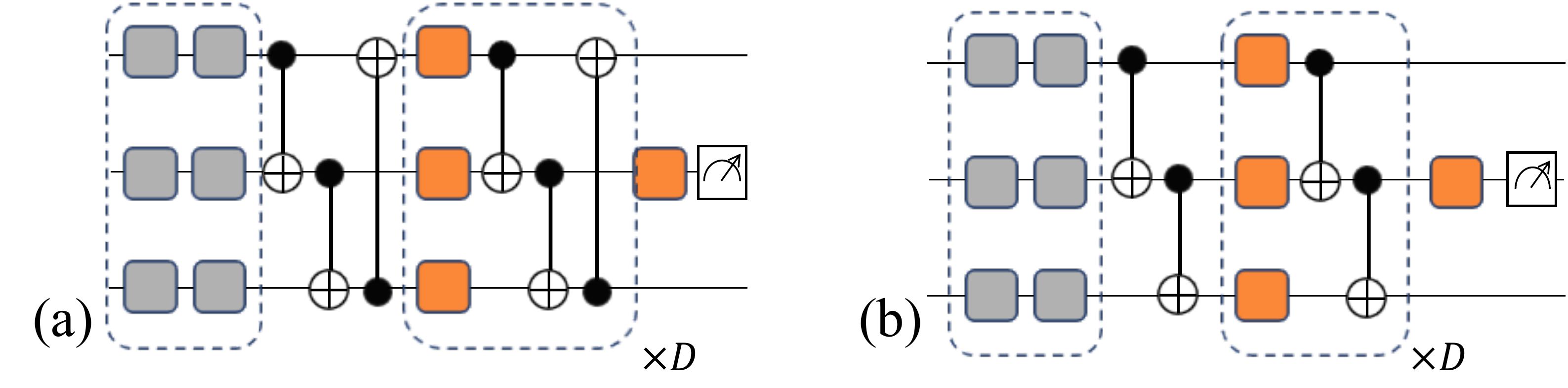}
  \caption{Schematic of the PQC trained as a regression model with different entangling layer layouts.  The encoding circuit block (grey boxes) is defined in \cite{mitarai2018quantum}.  The parameterized circuit block is constructed from RY (orange boxes) and CNOT gates.  (a) The ``cycle'' layout in which each parameterized block contains $3$ parameters and $3$ CNOT gates.  (b) The ``chain'' layout in which each parameterized block contains $3$ parameters and $2$ CNOT gates.  The final value predicted from the PQC is defined by the Pauli-Z expectation value measured on the second qubit.}
  \label{fig:qcl_circuit}
\end{figure}

A PQC can be described by a parameterized unitary operator $\mathcal{U}(\overrightarrow{x},\THETAVEC{})$ where $\overrightarrow{x}$ is an input feature and \THETAVEC{} is the P-dimensional vector of rotational parameters. Our study uses one-dimensional features, so we will replace $\overrightarrow{x}$ by $x$, a scalar. This operator can be decomposed as two operators $\mathcal{U}(x,\THETAVEC{})=\mathcal{U}_{\mathcal{E}}(x)\mathcal{U}_{P}(\THETAVEC{})$.  The first unitary $\mathcal{U}_{\mathcal{E}}(x)$ is the encoding circuit, which maps the input feature ($x$) into an initial state $|\psi(x)\rangle$.  We use the encoding circuit that was defined in \cite{mitarai2018quantum}: it is composed of $R_Y$ and $R_Z$ gates and contains no trainable parameters ($\mathcal{U}_{\mathcal{E}}(x)=\prod_
j R_Z^{(j)}(\cos^{-1}{(x^2)})R_Y^{(j)} (\sin^{-1}{(x)}$). The choice of this encoding restricts the training features to the domain $[-1,1]$.  The second unitary operator $\mathcal{U}_{P}(\THETAVEC{})$ is a parameterized circuit with single qubit rotations as trainable parameters. The specific circuit ansatzae that were used  for $\mathcal{U}_{P}(\THETAVEC{})$ are shown in Fig.~\ref{fig:qcl_circuit}.

\subsection{QCL}
\label{sec:qcl_overview}
The QCL workflow described in \cite{mitarai2018quantum} adapts  supervised learning to the variational training of PQCs. We use the QCL workflow to train a PQC as a simple regression model. The training of a regression model is done by encoding 1-dimensional features (x) into an initial state $|\psi(x)\rangle$, and the predicted label is the Pauli-Z expectation value of the second qubit in the register: $\hat{y} = \langle \psi(x) | \mathcal{M}| \psi(x) \rangle$ where $\mathcal{M} = \mathcal{U}_{P}(\THETAVEC{})^{\dagger}Z_1 \mathcal{U}_{P}(\THETAVEC{})$.  The choice of this label assignment restricts the predicted values to the range $[-1, 1]$.  $\mathcal{U}_{P}(\THETAVEC{})$ is parameterized by RY gates, and the number of parameters increases linearly with depth ($D$). The number of parameterized blocks in the circuit $P = |\overrightarrow{\mathbf{\theta}}| = 3D+1$.  

\begin{figure*}[htbp]
  \centering
  \includegraphics[width=\textwidth]{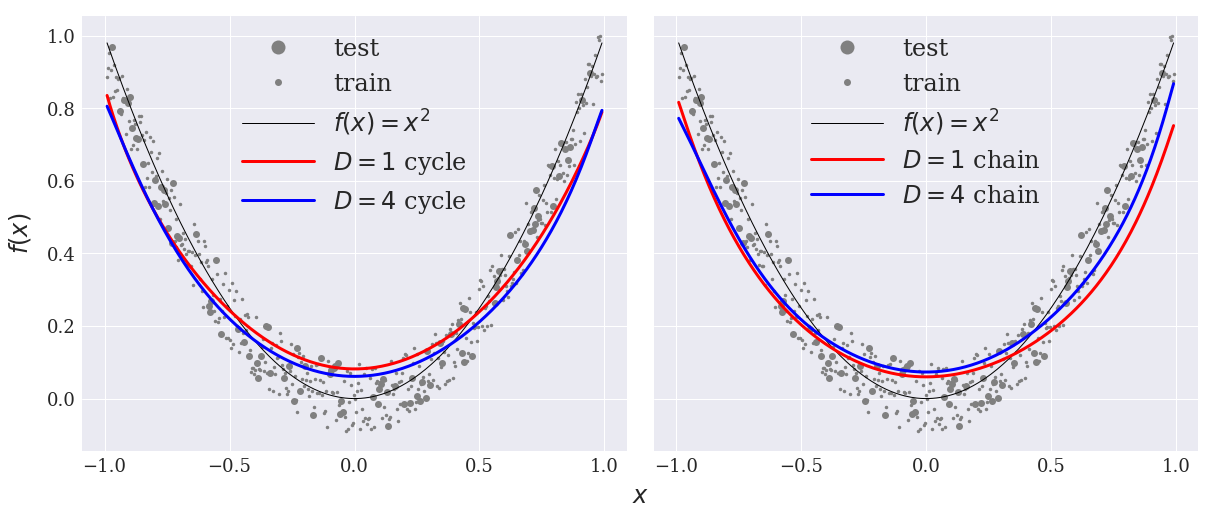}
  \caption{The quality of fits made by the smallest ($D=1$, 4 parameters) and largest circuits ($D=4$, 13 parameters) trained in this study. The fits are overlaid on the testing and training data.}
  \label{fig:quality_of_fit}
\end{figure*} 

A sample of $500$ points are drawn from a parabolic curve (second order polynomial) with added white noise.  The domain was constrained to $x \in [-1.0,1.0]$ and discretized into $500$ equally spaced points (one-dimensional features) which were labeled with the value $y = f(x) = x^2 + \epsilon $. The white noise factor was drawn uniformly from the interval $\epsilon \sim [-0.1, 0.1]$.  The $500$ points were divided into training and testing data using an 80/20 split (see Figure \ref{fig:quality_of_fit}). The loss function used to train our PQCs in Section \ref{sec:qcl} is the mean square error (MSE) evaluted over batches of $n$ samples drawn from the training set:
\begin{equation}
    \mathcal{L} = \frac{1}{n} \sum_{i=1}^n (y_i - \hat{y}_i)^2.
\label{eq:MSE_loss}
\end{equation}
The gradient of the loss function, 
\begin{equation}
    \nabla\mathcal{L} = -\frac{2}{n} \sum_{i=1}^n (y_i - \hat{y}_i)(\nabla \hat{y}_i)
\label{eq:loss_grad}
\end{equation}
is dependent on the circuit gradient which is evaluated by the parameter shift rule \cite{schuld2019evaluating}.

\subsection{Visualization methods}
\label{sec:visualization}
One challenge to understanding the loss landscape of PQCs is using appropriate methods to project the $P$-dimensional parameter space down onto $1$- or $2$- dimensions. Instead of varying along a single parameterized gate direction, we use parameter sets that update multiple components in the parameter vector. This technique has been used in classical machine learning for one-dimensional cuts \cite{goodfellow2014qualitatively,li2017visualizing}, and two dimensional projections \cite{goodfellow2014qualitatively,garipov2018loss}.  The use of interpolated directions distinguishes our work from other recent studies which use visualizations of the PQC loss landscape made by varying individual parameter directions \cite{huembeli2021characterizing,mari2021gradient,arrasmith2021equivalence}. The use of interpolated directions helps avoid bias in the loss landscape to frozen parameter subspaces that may not be indicative of global features. 

One-dimensional cuts of the PQC landscape are made using the method described in \cite{goodfellow2014qualitatively}.  Given a pair of $P$-dimensional vectors \THETAVEC{A} and \THETAVEC{B} in the loss landscape, there is a straight line that connects them and any vector that lies on this straight line is given by
\begin{align}
    \overrightarrow{\pi}(\alpha) = (1 - \alpha) \THETAVEC{A} + \alpha \THETAVEC{B} && \text{for}\: \alpha \in [0,1]. 
\end{align}

All of the circuits in this study are trained in parameter spaces $\mathbb{R}^P$ with $\dim{\mathbb{R}^P}>2$.  Contour plots (two-dimensional cuts) are defined by projecting points in the $P$-dimensional loss landscape onto a $2$-dimensional plane defined by (\THETAVEC{A},\THETAVEC{B},\THETAVEC{C}). One point (\THETAVEC{A}) is chosen to define the origin, then the two orthogonal directions in the plane are defined as: $w_1 = \THETAVEC{A}-\THETAVEC{B}$ and the second direction $w_2$ (orthogonal to $w_1$), is found using the Gram-Schmidt method \cite{luenberger1997optimization}, and fixed such that \THETAVEC{C} sits on the $w_2$ axis. Both directions are normalized and \texttt{numpy.meshgrid} is used to generate a grid of points in this plane in which the MSE loss is evaluated.

\subsection{Nudged elastic band}
\label{sec:neb}
We also use interpolated parameter sets to analyze the mode connectivity in the loss landscape.  In Section \ref{sec:dicussion} we show how different minima can be connected in the loss landscape using the nudged elastic band algorithm (NEB) \cite{henkelman2000neb,draxler2018essentially}.  This method has been applied to the study of mode connectivity in classical deep learning models \cite{draxler2018essentially}, and we use a modified version of the improved tangent method~\cite{henkelman2000neb} to find low loss paths in PQC landscapes.  

NEB initializes a set of ($t$) pivot points $\overrightarrow{\pi}(t)$ along a straight line path connecting two minima (\THETAVEC{A}, \THETAVEC{B}). The endpoints are fixed ($\overrightarrow{\pi}(0)=\THETAVEC{A}$, $\overrightarrow{\pi}(t-1)=\THETAVEC{B}$), but the location of the interior pivots are shifted in order to minimize the loss along the path.  Between neighboring pivot points there is a tangent vector defined along $\THETAVEC{i+1}-\THETAVEC{i}$. The pivot points are updated in the $P$-dimensional space, and the update to each pivot location is defined by $2$ terms:   a spring force that acts parallel to the tangent vector (parameterized by a spring constant $k$), and a force that is directed perpendicular to this tangent vector and is parameterized by $\nabla \mathcal{L}$.  When applying NEB to our QCL landscape, we do not have an explicit potential function describing the landscape, but we do have a finite set of data points.  To reduce computational overhead we use batches of training samples to estimate the loss value $\tilde{\mathcal{L}}_i$ at a pivot point \THETAVEC{i}, and batches of training samples to estimate the MSE loss gradient $\tilde{\nabla{\mathcal{L}}}_i$ (which is again computed using the parameter shift rule \cite{schuld2019evaluating}).  

While both NEB and 1D cuts use interpolated straight line paths $\overrightarrow{\pi}$, we parameterize NEB paths with the pivot index  ($\overrightarrow{\pi}(t), t \in [0,n-1]$). Interpolated paths which are single straight lines are parameterized with $\overrightarrow{\pi}(\alpha), \alpha \in [0,1]$. With a small number of steps and pivot points, we effectively limit to a small search around a straight line segment, we call this localized NEB. We also allow for more exhaustive searches that result in piece-wise paths.

\section{QCL (supervised learning)}\label{sec:qcl}

We implement our QCL workflow in Pennylane \cite{bergholm2020pennylane} which gave us the flexibility to switch between $3$ different gradient-based optimizers: standard stochastic gradient descent (SGD) with a fixed learning rate \cite{goodfellow2016deeplearning}, the Adam stochastic optimizer \cite{kingma2014adam}, and the quantum natural gradient (QNG) \cite{stokes2020quantum} on quantum circuit learning tasks.  We fix the total number of optimization steps at $300$, do not implement early stopping, and choose a learning rate of $\alpha = 0.05$ for all $3$ methods.

We trained PQCs with $D = 1, 2, 3, 4$ (shown in Fig.~\ref{fig:qcl_circuit}).  For each combination of depth, layout and optimizer we ran $36$ training runs for $300$ steps. Five training runs used an adapted version of Glorot intialization \cite{glorot2010understanding}. Since the entangling layers entangle all qubits, we replaced the fan-in/fan-out dependence with the number of qubits in the PQC.  Thus the initial values: $\overrightarrow{\theta_0}$ are drawn from Gaussians with standard deviation ($\sigma = \sqrt{2/6}$) centered at either $\mu = \lbrace 0, \pi/4, \pi/2, 3\pi/4,\pi\rbrace$. We also include an initialization drawn uniformly from the interval $[0, 2\pi)$.  For each initialization we trained the circuits with batch sizes $\lbrace 1,2,4,8,16,32\rbrace$.  We note that for the online learning (batch size 1) case, $300$ steps of optimization will not use all the training data to train the PQC.

\begin{figure*}[htbp]
  \centering
  \includegraphics[width=\textwidth]{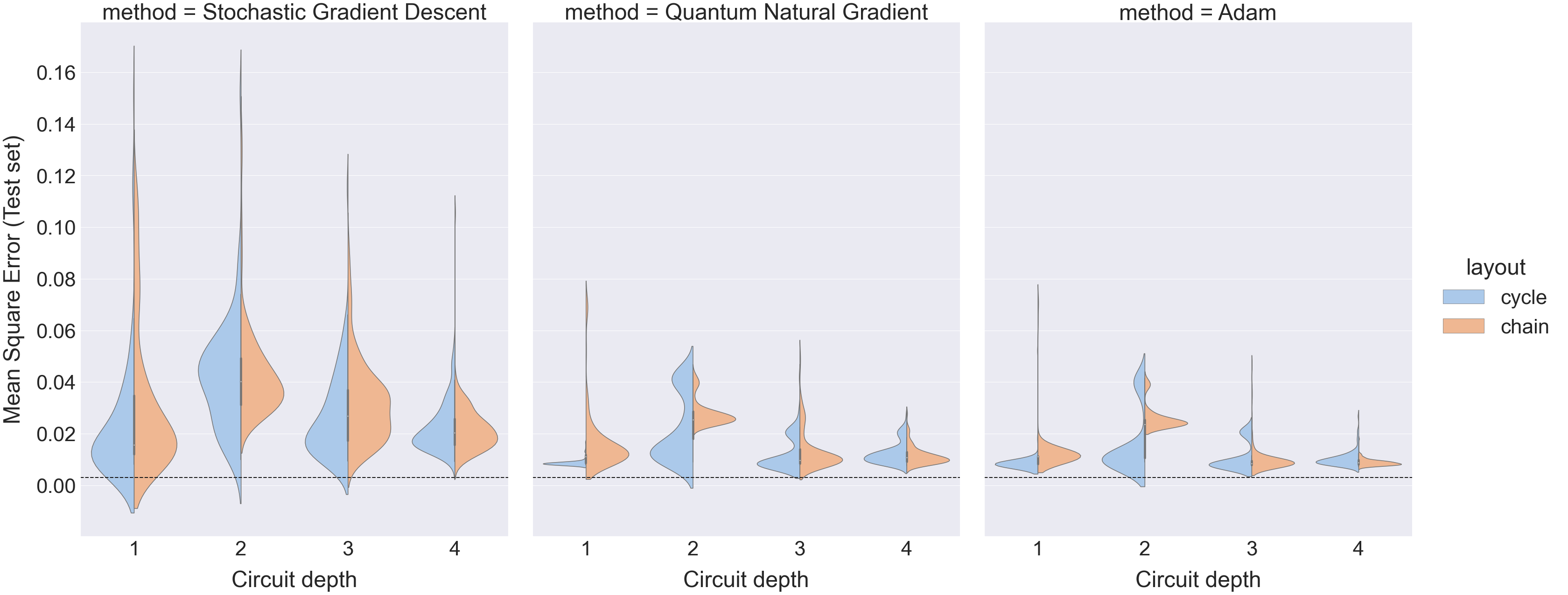}
  \caption{The minimum test set MSE for multiple training runs of PQCs with circuit depth $D\leq4$. Horizontal guide line (black, dashed) included to highlight MSE lower bound.}
  \label{fig:violin_plot}
\end{figure*} 
In Figure \ref{fig:violin_plot} we show the distribution of MSE test values for all trained PQCs, plotted as a function of entangling layer design, depth and optimizer choice.  We define a lower bound for the MSE using the noiseless curve.  If a PQC could predict values that lie exactly along the noiseless curve $f(x)=x^2$, the test set MSE would be $0.003$.  While the distributions in Fig. \ref{fig:violin_plot} may have tails that dip below this threshold, this is an artifact of the kernel density estimate fit -- in our observed results there were no circuits that returned a value of $\mathrm{MSE}\leq 0.003$ on the test data.  

For each distribution plotted in Fig. \ref{fig:violin_plot}, we report the median value of the MSE distribution in Table \ref{tab:mse_values}. However, there are several instances of strong secondary modes in the MSE test distribution. To count the number of training runs that converge to low loss regions, we generate a histogram over the range of MSE values $[0.007,0.16]$ using 50 equal width bins. In Table \ref{tab:bin_values} we report the midpoint of the lowest occupied bin and the number of training runs in that bin. 
\begin{table}[t]
  \centering
 \caption{Median of the MSE test distribution for circuits with different entangling layouts, optimizer methods and parameters. The number of individual training runs used to compute each statistical property is (n).}
 
\begin{tabular}{|c|c|c|c|c|c|c|}
 \hline
Layout &  Method & Depth & Median & n\\
 \hline
 \hline
 \multirow{12}*{Cycle}
            & \multirow{4}*{SGD}
&1  &  0.0126 &  108 \\
&&2  &  0.0426 &   90 \\
&&3 &  0.0213 &  108 \\
&&4 &  0.0201 &  108 \\
\cline{2-5}
            & \multirow{4}{*}{QNG}
&1  &  0.0085 &  114 \\
&&2  &  0.0156 &  108 \\
&&3 &  0.0085 &  108 \\
&&4 &  0.0114 &  144 \\
\cline{2-5}
            & \multirow{4}{*}{Adam}
&1  &  0.0081 &  108 \\
&&2  &  0.0099 &  108 \\
&&3 &  0.0080 &  102 \\
&&4 &  0.0093 &  108 \\
\cline{2-5}
\hline
\hline
 \multirow{12}*{Chain}
            & \multirow{4}*{SGD}
&1  &  0.0181 &  180 \\
&&2  &  0.0386 &  150 \\
&&3 &  0.0310 &  180 \\
&&4 &  0.0207 &  174 \\
\cline{2-5}
            & \multirow{4}*{QNG}
&1  &  0.0115 &  144 \\
&&2  &  0.0262 &  144 \\
&&3 &  0.0103 &  144 \\
&&4 &  0.0104 &  144 \\
\cline{2-5}
            & \multirow{4}*{Adam}
&1  &  0.0113 &  144 \\
&&2  &  0.0242 &  144 \\
&&3 &  0.0088 &  144 \\
&&4 &  0.0085 &  144 \\
\hline
\end{tabular}
\label{tab:mse_values}
\end{table}
\begin{table}[t]
  \centering
 \caption{Midpoint of the lowest occupied bin of the MSE test histogram, the number of training runs which reach this bin ($n_{occ}$) and mean number of steps to get to the lowest MSE test value ($\langle n_s \rangle$) for circuits with different entangling layouts, optimizer methods and rotation gates.}
 
\begin{tabular}{|c|c|c|c|c|c|c|}
 \hline
Layout &  Method & Depth & Lowest Bin & $n_{occ}$ & $\langle n_s \rangle$\\
 \hline
 \hline
 \multirow{12}*{Cycle}
            & \multirow{4}*{SGD}
&1  &  0.008 &  31 & 296\\
&&2  &  0.008 &   1 & 290\\
&&3 &  0.008 &  12 & 232\\
&&4 &  0.008 &   1 & 300\\
\cline{2-6}
            & \multirow{4}{*}{QNG}
&1  &  0.008 &  108 & 267\\
&&2  &  0.008 &   10 & 250\\
&&3 &  0.008 &   73 & 282\\
&&4 &  0.008 &   50 & 279\\
\cline{2-6}
            & \multirow{4}{*}{Adam}
&1  &  0.008 &  105 & 237\\
&&2  &  0.008 &   56 & 248\\
&&3 &  0.008 &   78 & 232\\
&&4 &  0.008 &   68 & 257\\
\cline{2-6}
\hline
\hline
 \multirow{12}*{Chain}
            & \multirow{4}*{SGD}
&1  &  0.0116 &  47 & 285\\
&&2  &  0.0269 &   9 & 290\\
&&3 &  0.0116 &  10 & 295\\
&&4 &  0.0080 &   2 & 293\\
\cline{2-6}
            & \multirow{4}*{QNG}
&1  &  0.0116 &  98 & 227\\
&&2  &  0.0238 &  51 & 286\\
&&3 &  0.0080 &  71 & 282\\
&&4 &  0.0080 &  60 & 283\\
\cline{2-6}
            & \multirow{4}*{Adam}
&1  &  0.0116 &  134 & 217\\
&&2  &  0.0238 &  103 & 266\\
&&3 &  0.0080 &  122 & 247\\
&&4 &  0.0080 &  129 & 256\\
\hline
\end{tabular}
\label{tab:bin_values}
\end{table}
Figure \ref{fig:violin_plot} and Table \ref{tab:bin_values} show that a large percentage of the QNG and Adam training runs converge to very low loss values for both the chain and cycle layouts.  

\section{Mode connectivity}
\label{sec:connectivity}

Mode connectivity has been qualitatively studied in the classical machine learning literature \cite{draxler2018essentially,garipov2018loss,kuditipudi2019connectivity}, and we agree with the following observation for gradient-based training:  connectivity of low loss regions in the training landscape does not guarantee that any gradient-based training will converge to a low loss region.  The extreme outliers large MSE shown in Fig.~\ref{fig:violin_plot}  are examples of gradient-based training becoming trapped. 

In the loss landscape there are multiple minima created by rotational symmetry of the parameterized gates \cite{huembeli2021characterizing}, and in overparameterized models there is exponential growth in the number of loss landscape minima.  We empirically confirm this by observing that the training runs which result in low MSE loss on the test set do not converge to the same parameter set, or even the same region of parameter space. Table \ref{tab:bin_values} shows that many training runs converge to regions of low MSE loss -- we reduce the number of unique parameter sets using mean shift clustering implemented in \texttt{scikit-learn} \cite{scikit-learn} and identify parameter sets that correspond to unique minima of the landscape from the mean shift cluster centers. We detail the clustering results in Appendix \ref{appendix:clustering} and refer to these unique regions of low loss as the aggregated minima set (AMS).

NEB analysis can be used to find connected paths of low loss \cite{draxler2018essentially,garipov2018loss}, but it can also be used to characterize saddle points in the landscape \cite{henkelman2000neb,wales1998archetypal,draxler2018essentially}. Here we use the modified NEB algorithm described in Section~\ref{sec:neb} to find connected regions of the loss landscape and show that elements of the AMS can be connected by paths where the loss minimally changes. We first define connectivity in our landscapes: in Section \ref{sec:qcl_overview}, we sorted the test MSE values into discrete levels, using a histogram of $50$ bins and the lowest bin occupied was used to compare training success.  In other studies of loss landscape connectivity \cite{draxler2018essentially,abbas2020power,garipov2018loss}, the connectivity between two minima is described qualitatively as ``training loss remains essentially at the same value as at the minima'' \cite{draxler2018essentially} or that the ``train loss and test error remain low along these paths'' \cite{garipov2018loss}.  

In this section, we will quantify the connectedness between two minima based on the highest point along the path $\max{\mathcal{L}(\pi(t))}$. We compare how large is $\max{\mathcal{L}(\pi(t))}$ relative to the endpoints (the minima at \THETAVEC{A} and \THETAVEC{B}) and how much $\max{\mathcal{L}(\pi(t))}$ can be reduced by updating the NEB pivot points.

We start by searching for a specific feature in the loss landscape: ravines. These are large connected regions of low MSE loss that appear to be separated by high barriers.  They are also found in the loss landscape of over-parameterized neural networks and have played an important role in understanding the success of deep learning \cite{hochreiter1995simplifying,hochreiter1997flat,baldassi2020shaping,kuditipudi2019connectivity}. 

To search for ravines we run batch gradient NEB on all pairs of minima in the AMS using a ``localized'' search.  This is not an exhaustive search in the loss landscape; with a low number of pivot points ($10$), a small number of update steps ($10$), and a small learning rate $0.05$ we are searching over paths that are very close to a single straight line.  We use this local search to find pairs of minima that are connected by straight line paths (or nearly straight lines).  If two points sit in the same ravine, then NEB will quickly converge to a path that has low loss value at each pivot point. 

\begin{figure}[htbp]
  \centering
  \includegraphics[width=\columnwidth]{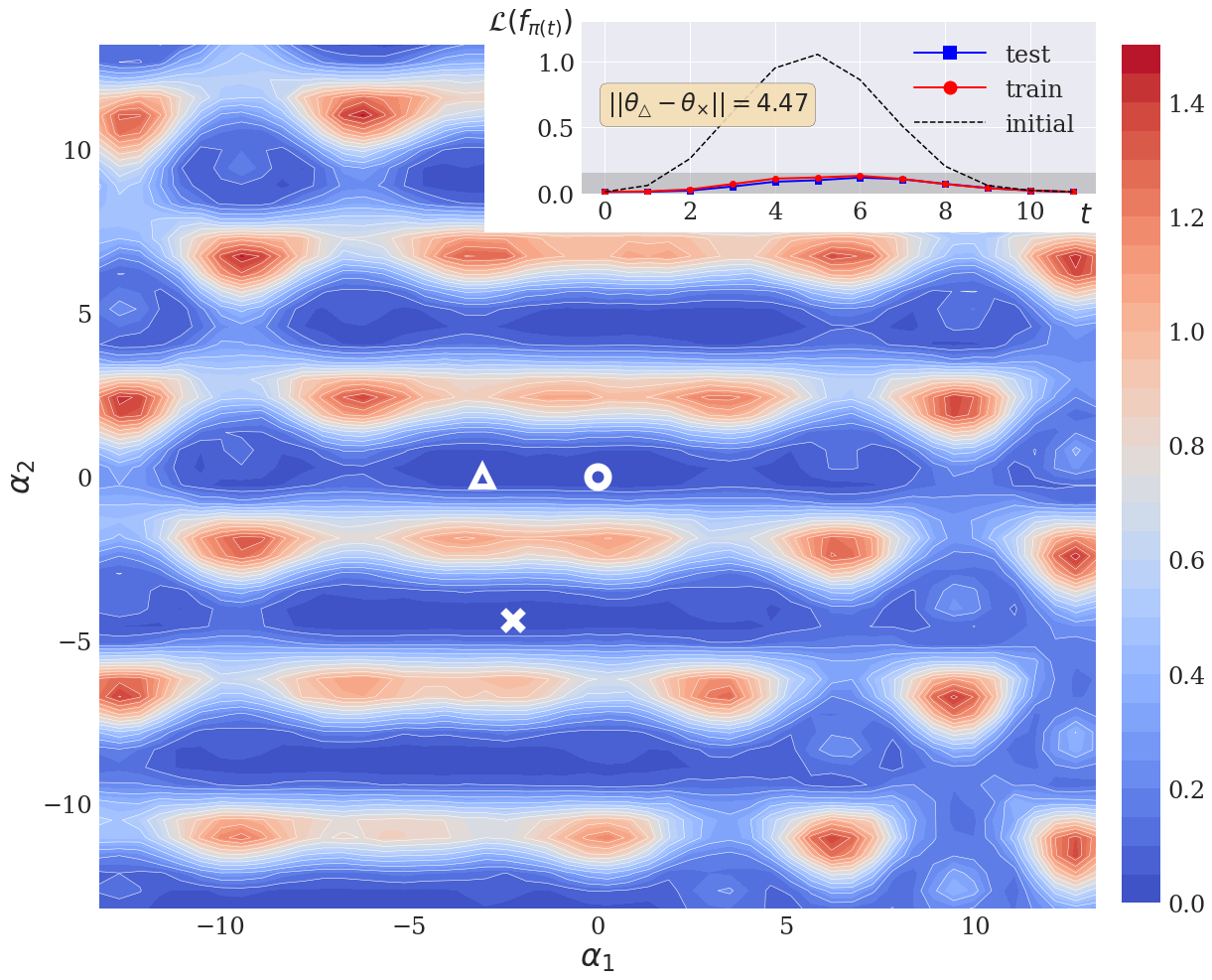}
  \caption{Examples of ravines in the loss landscape for ($D=1$) circuits with the chain layout.  The axes are defined by the interpolated directions $w_1,w_2$.  (Inset) The change in loss along a piece-wise linear path connecting two minima in separate ravines. Grey shaded region shows the range of values plotted in Fig. \ref{fig:violin_plot}.}
  \label{fig:4P_chain_landscape}
\end{figure}

\begin{figure}[htbp]
  \centering
  \includegraphics[width=\columnwidth]{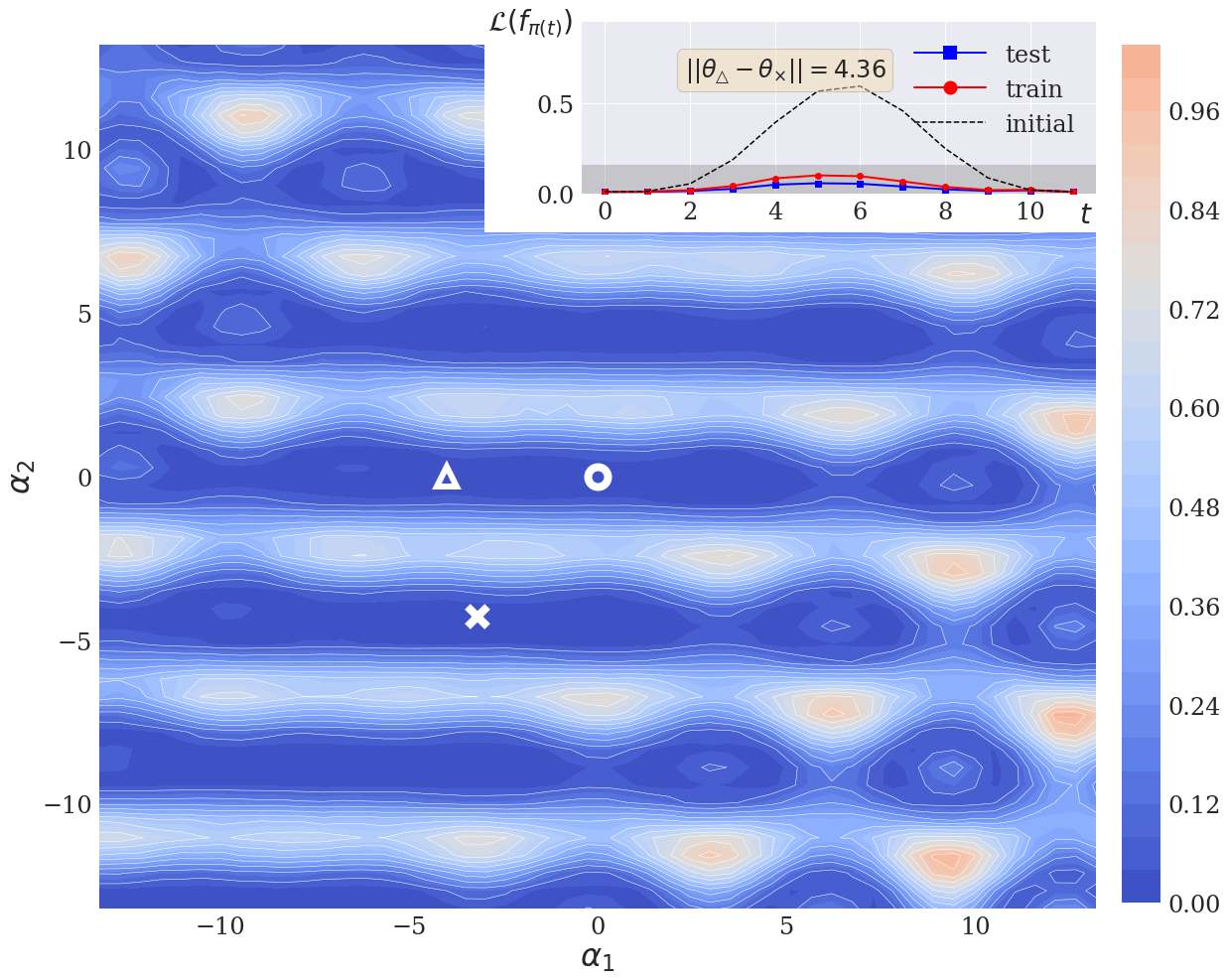}
  \caption{Examples of ravines in the loss landscape for ($D=1$) circuits with the cycle layout.  The axes are defined by the interpolated directions $w_1,w_2$.  (Inset) The change in loss along a piece-wise linear path connecting two minima in separate ravines. Grey shaded region shows the range of values plotted in Fig. \ref{fig:violin_plot}.}
  \label{fig:4P_cycle_landscape}
\end{figure}

\begin{figure}[htbp]
  \centering
  \includegraphics[width=\columnwidth]{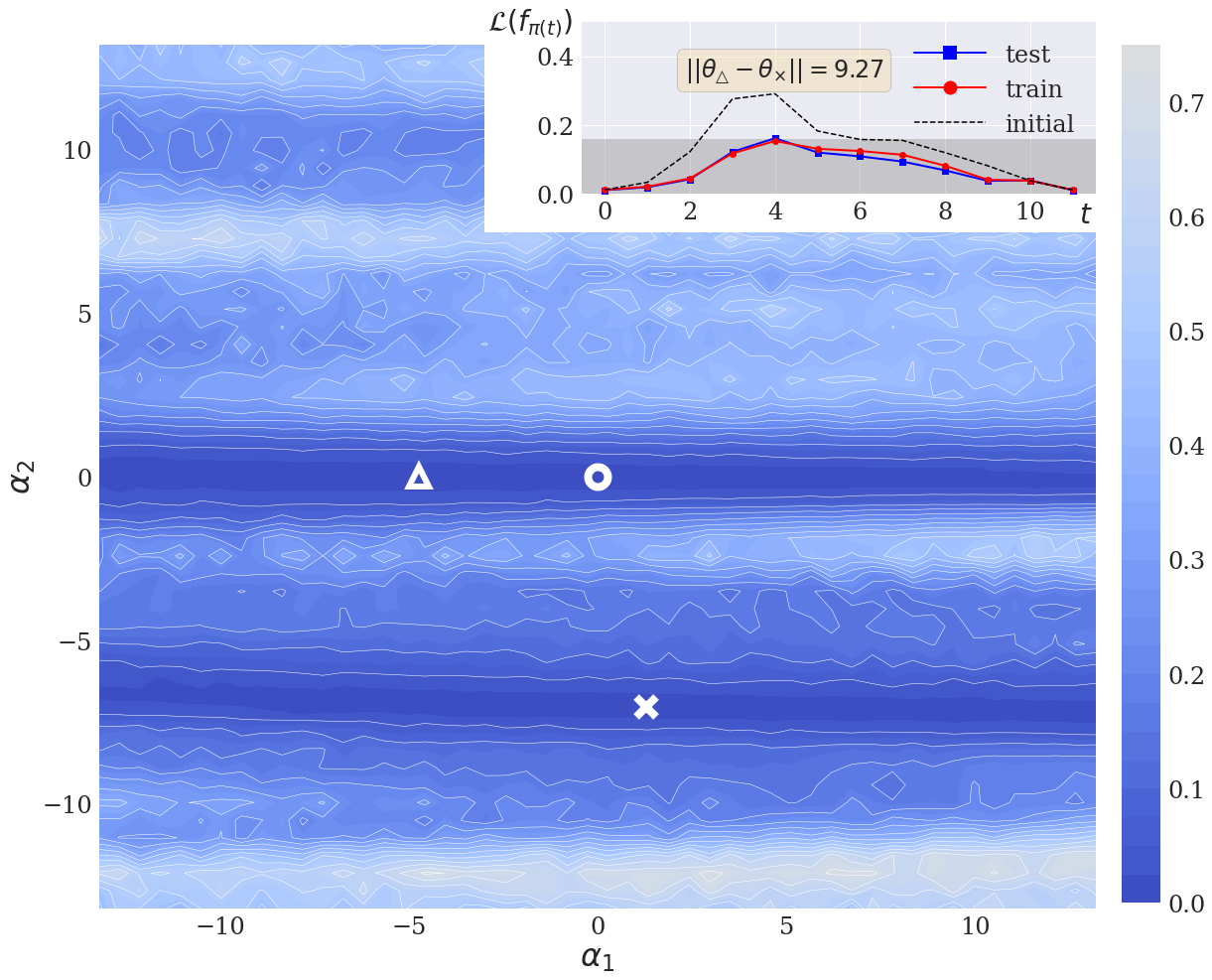}
  \caption{Ravines found in the loss landscape for ($D=2$) circuits with the cycle layout. The axes are defined by the interpolated directions $w_1,w_2$.  (Inset) The change in loss along a piece-wise linear path connecting two minima in separate ravines. Grey shaded region shows the range of values plotted in Fig. \ref{fig:violin_plot}.}
  \label{fig:7P_cycle_landscape}
\end{figure}

\begin{figure}[htbp]
  \centering
  \includegraphics[width=\columnwidth]{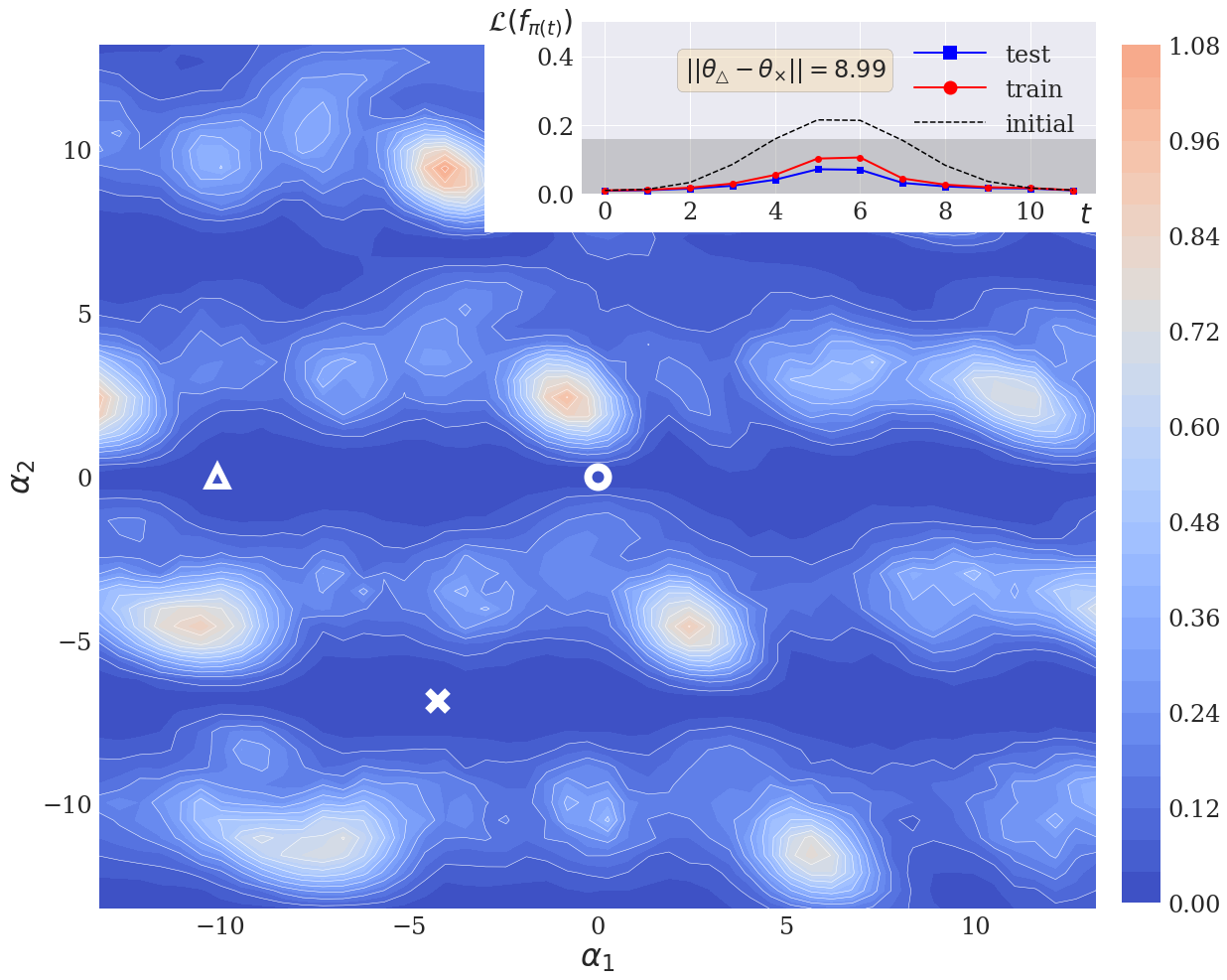}
  \caption{Ravines found in the loss landscape for ($D=3$) circuits with the cycle layout. The axes are defined by the interpolated directions $w_1,w_2$.  (Inset) The change in loss along a piece-wise linear path connecting two minima in separate ravines. Grey shaded region shows the range of values plotted in Fig. \ref{fig:violin_plot}.}
  \label{fig:10P_cycle_landscape}
\end{figure}

Using the AMS and NEB, we identify connected regions in the $D=1,2,3$ landscapes for circuits with cycle layouts; and the $D=1$ landscape for circuits with the chain layout.  To visualize these landscapes we choose $3$ minima (\THETAVEC{A},\THETAVEC{B},\THETAVEC{C}) in the respective AMS, and generate the contour plot c.~f.~Section~\ref{sec:visualization}. \THETAVEC{A} and \THETAVEC{B} are minima connected by a straight line of low loss and are assumed to sit in a ravine, while \THETAVEC{C} is a point that is not connected to either \THETAVEC{A},\THETAVEC{B} by a straight line of low loss. The landscape plots shown in Figs. \ref{fig:4P_chain_landscape} - \ref{fig:10P_cycle_landscape} verify what we predict through NEB analysis: large regions of low loss exist in the landscape.  

\begin{figure}[htbp]
  \centering
  \includegraphics[width=\columnwidth]{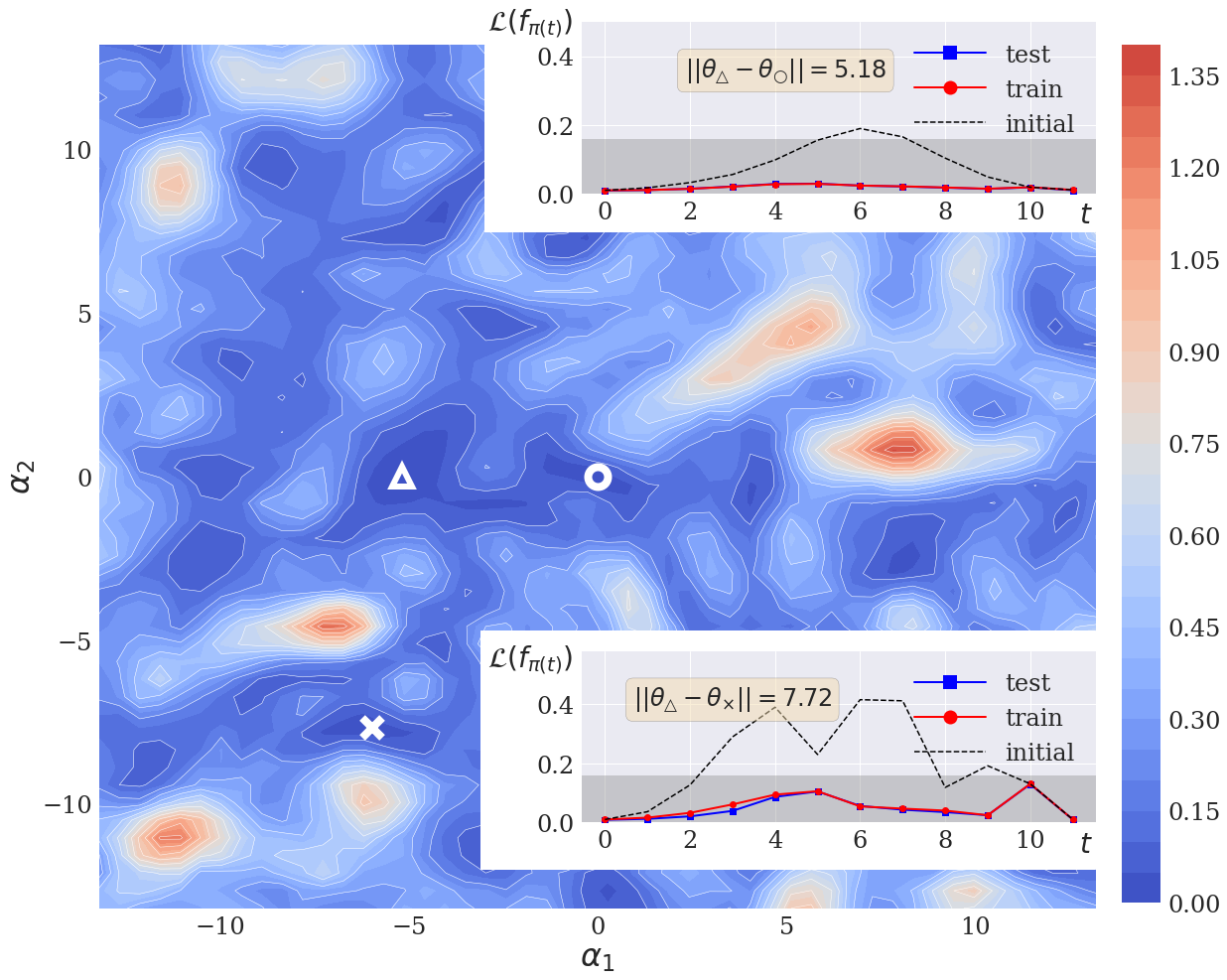}
  \caption{Ravines found in the loss landscape for ($D=4$) circuits with the cycle layout. The axes are defined by the interpolated directions $w_1,w_2$.  (Inset) The change in loss along a piece-wise linear path connecting two minima in separate ravines. Grey shaded region shows the range of values plotted in Fig. \ref{fig:violin_plot}.}
  \label{fig:13P_cycle_landscape}
\end{figure}

For our second application of NEB, we search for a path of low loss connecting two points in adjacent ravines. Instead of using a straight line of interpolated values between these points (which would travel a path that climbs up and over the large barriers between the ravines), we use NEB with 12 pivot points (10 of which are updated) and $100$ update steps to find an alternate path in the landscape.  The results are shown in the inset plots of Figs. \ref{fig:4P_chain_landscape} - \ref{fig:10P_cycle_landscape}. The piece-wise paths are not as flat as in the ravines, but NEB was able to find piece-wise paths with lower loss values than the straight line connections. In Table~\ref{tab:loss_values_max} we show that maximum loss value evaluating along the initial band ($\pi_0$) and along the best NEB paths ($\pi^{\prime}$). We re-iterate that the pivot positions are updated in the full parameter space, and the best NEB path is not constrained to the plane used to define a contour plot.  For the initial band we only use the training data, but with the best NEB paths we show the maximum loss values for the training and testing data sets. We also report the largest ratio of the loss value along NEB with respect to the endpoints $\max{(\max{\mathcal{L}_{\pi^{\prime}}}/\mathcal{L}_{\protect\THETAVEC{}})}$.
\begin{table*}[t]
  \centering
 \caption{Maximum loss values evaluated along the initial and best NEB paths, shown in the inset plots of Figs. \ref{fig:4P_chain_landscape} - \ref{fig:10P_cycle_landscape}. The largest ratio $\max{\mathcal{L}_{\pi^{\prime}}}$ and the NEB endpoints is also given.}
 
\begin{tabular}{|c| c| c| c| c| c| c| c|}
\cline{5-8}
\multicolumn{4}{c}{} & \multicolumn{2}{|c|}{train}&\multicolumn{2}{|c|}{test} \\
\hline
 Depth &  Layout & Markers & $\max{\mathcal{L}_{\pi(0)}}$   &$\max{\mathcal{L}_{\pi^{\prime}}}$   & $\max{(\max{\mathcal{L}_{\pi^{\prime}}}/\mathcal{L}_{\protect\THETAVEC{}})}$ & $\max{\mathcal{L}_{\pi^{\prime}}}$  & $\max{(\max{\mathcal{L}_{\pi^{\prime}}}/\mathcal{L}_{\protect\THETAVEC{}})}$ \\
 \hline
1  & chain & $\bigtriangleup \to \times$ & 1.054 & 0.133  & 11.2 & 0.120 & 10.7\\
1  & cycle  & $\bigtriangleup \to \times$ & 0.594 & 0.099 & 11.3 & 0.056 & 6.86  \\
2  & cycle  & $\bigtriangleup \to \times$ & 0.290 & 0.153 & 14.8 & 0.161 & 17.1 \\
3 & cycle  & $\bigtriangleup \to \times$ & 0.214 & 0.104 & 11.9 & 0.070 & 8.0 \\
\hline
\hline
4 & cycle  & $\bigtriangleup \to \bigcirc$ & 0.189 & 0.028 & 3.1 & 0.028 & 3.17\\
4  & cycle  & $\bigtriangleup \to \times$ & 0.415 & 0.131 & 12.8 & 0.130 & 12.9 \\

\hline
\end{tabular}
\label{tab:loss_values_max}
\end{table*}
For most ravines, the algorithm is able to find paths where there is a significant reduction in the maximum loss along the NEB. While the maximum loss value may be nearly $20$ times larger than the endpoint values, the loss values along the NEB remain in the range of values plotted in Fig. \ref{fig:violin_plot}, with the only exception being the value of $\max{\mathcal{L}_{\pi^{\prime}}}$ evaluated on the test set for the $D=2$ circuit. Using the same local NEB search over all pairs of minima, no ravines were found in $D=4$ circuits.  However, for $D=4$ cycle circuits, the longer NEB search could find low loss paths connecting two minima (markers $\bigtriangleup \to \bigcirc$ in Fig.~\ref{fig:13P_cycle_landscape}).

\subsection{Ravines and narrow gorges}
The loss function (or cost function) landscape has been studied in terms of identifying barren plateaus \cite{mcclean2018barren} and their sources \cite{cerezo2021cost,wang2020noiseinduced,marrero2020entanglement}.  In addition to barren plateaus, the concept of ``narrow gorges,'' has been introduced in \cite{cerezo2021cost,arrasmith2021equivalence}.  Narrow gorges are characterized by a contraction of the basin of attraction around the global optima in the loss landscape, and may be accompanied by a flattening of the landscape, but they are different from ravines in several key ways.  

First, the learning tasks employed in \cite{cerezo2021cost,arrasmith2021equivalence} studied variational quantum algorithms which take the fixed initial state $|0\rangle$ to a fixed target state $|\psi \rangle$, where the cost function is dependent on the distance between a prepared state and the target state.  The QCL training using here relies on training PQCs that define a label from feature-dependent input states $|\psi(x)\rangle$.  The impact of feature embedding on the loss landscape and its connection to circuit expressibility is an active area of study~\cite{havlivcek2019supervised,lloyd2020quantum,perez2020data,schuld2021effect}.  Second,while we study smaller widths and depths, we see that the existence of ravines does not necessarily mean the landscape will be flat. For some landscapes (see Fig. \ref{fig:7P_cycle_landscape}) the ravines are separated by low barriers that are difficult to navigate (NEB does not significantly reduce the value of $\max{\mathcal{L}_{\pi}}$), while for other landscapes (see Fig. \ref{fig:4P_cycle_landscape}) the ravines are separated by high barriers but low loss paths exist between them.

\subsection{General Piece-wise path connectivity}
\label{sec:piecewise}
The previous section focused on one specific landscape feature (ravines), which are not required for successful training. For circuits with chain layouts and $D>1$, we do not find any large ravines, but the gradient-based training is able to train these circuits to relatively low loss. It is possible that two minima in the AMS that can be connected by a path of piece-wise linear paths (instead of a single line segment). For example, consider $D=4$ circuits with cycle layouts: of the $306$ total training runs, $119$ found circuits with low test loss, yet the local NEB analysis did not uncover any ravines. 

\begin{figure}[htbp]
  \centering
  \includegraphics[width=\columnwidth]{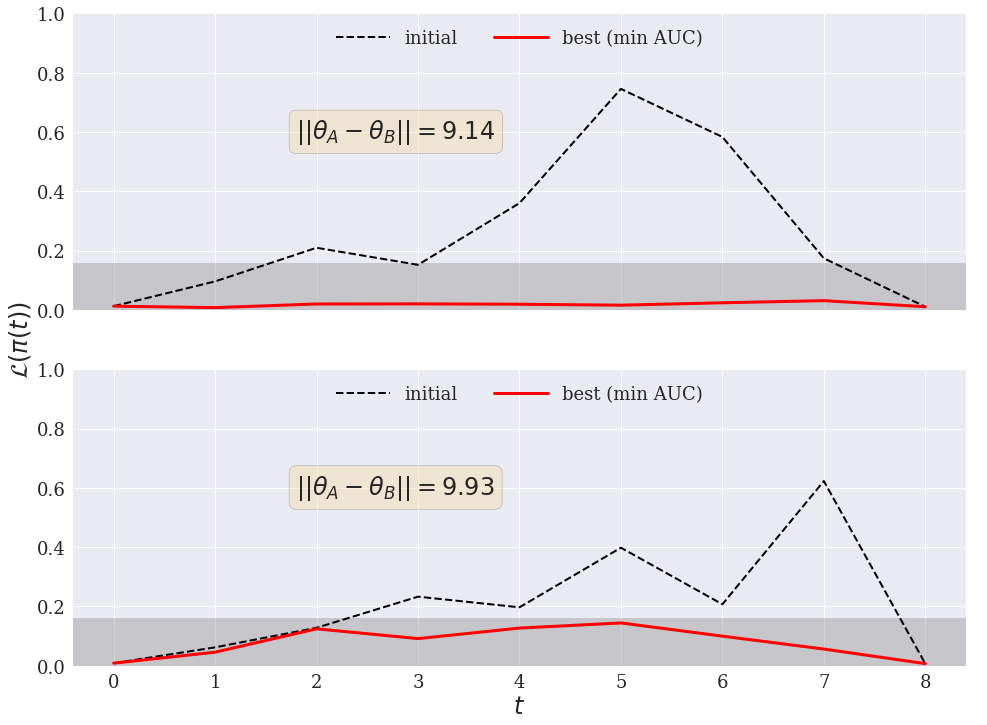}
  \caption{With $50$ steps of pivot updates we find minima that are nearly connected $\epsilon \to 0$ with the $D=4$ cycle layout circuits (top panel). We also show an example of minima which are not connected $\epsilon \neq 0$.  The NEB had 9 pivot points, 7 of which can be updated.  The Euclidean distances between the fixed endpoints are also shown $\lVert \protect\THETAVEC{A} - \protect\THETAVEC{B} \rVert $. Grey shaded region shows the range of values plotted in Fig. \ref{fig:violin_plot}.}
  \label{fig:D4_cycle_NEB_50pts}
\end{figure} 

We next use the NEB analysis to find piece-wise linear paths of low loss.  
We observe that piece-wise paths of low loss can indeed connect these local minima.  Based on the low-loss curve uncovered by the NEB analysis, this indicates that mode-inter-connectivity can persist for deeper circuits.  However, we emphasize that any two minima are not guaranteed to be connected by a low loss path. 

\begin{figure}[htbp]
  \centering
  \includegraphics[width=\columnwidth]{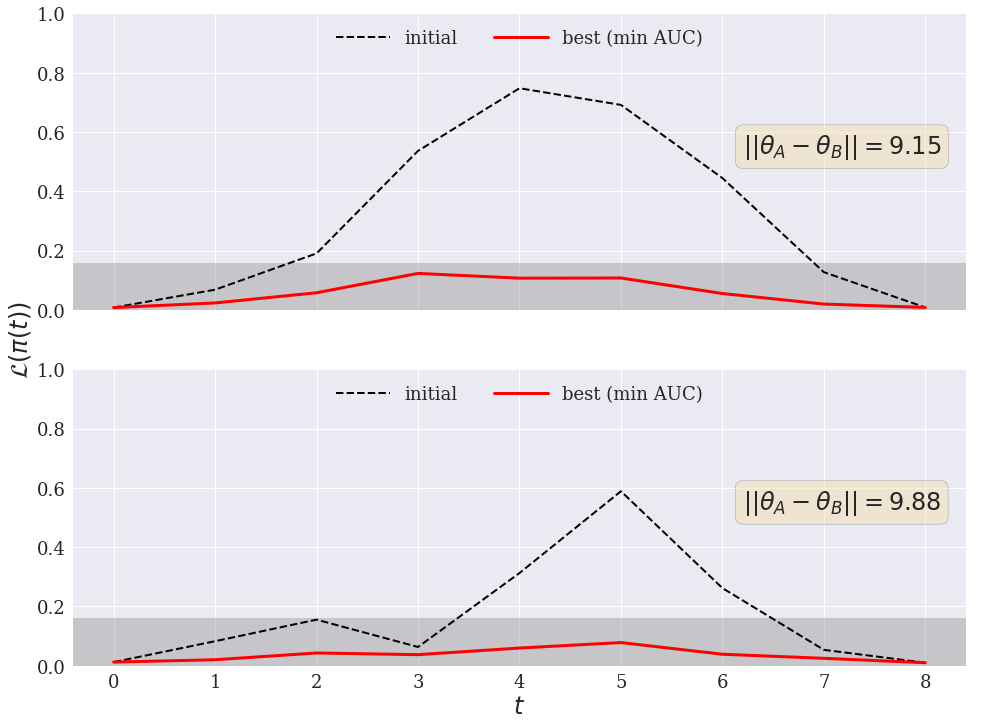}
  \caption{With $50$ steps of pivot updates we find minima that are nearly connected $\epsilon \to 0$ with $D=4$ the chain layout circuits (top panel). We also show an example of minima which are not connected $\epsilon \neq 0$.  The NEB had 9 pivot points, 7 of which can be updated.  The Euclidean distances between the fixed endpoints are also shown $\lVert \protect\THETAVEC{A} - \protect\THETAVEC{B} \rVert $. Grey shaded region shows the range of values plotted in Fig. \ref{fig:violin_plot}.}
  \label{fig:D4_chain_NEB_50pts}
\end{figure}

For ($D = 4$) circuits (both chain and cycle layouts), we chose two pairs of minima from each AMS, then we re-ran the local NEB analysis but increased the number of pivot updates from $10$ to $50$.  We chose sets of minima that were separated by similar Euclidean distances.  In Figs. \ref{fig:D4_cycle_NEB_50pts}, \ref{fig:D4_chain_NEB_50pts} we plot $\mathcal{L}_{\pi(t)}$, the loss evaluated along each NEB curve.  The initial curve is the straight line cut between two minima, and we also plot the best NEB curve, quantified by the minimum area under curve (AUC). 

For the cycle layout: for one set of minima (Fig. \ref{fig:D4_cycle_NEB_50pts} top panel) we found a piece-wise linear path which reduced the AUC from $2.0578\to 0.379$ and reduced the maximum loss value along each NEB curve from $0.745 \to 0.0315$ with $\max{(\max{\mathcal{L}_{\pi^{\prime}}}/\mathcal{L}_{\protect\THETAVEC{}})}=2.88$.  For the second set of minima (Fig. \ref{fig:D4_cycle_NEB_50pts} (bottom panel)), the AUC was reduced from $2.706 \to 0.7798$ and the maximum loss value for the NEB curve was reduced from $0.623 \to 0.145$ with $\max{(\max{\mathcal{L}_{\pi^{\prime}}}/\mathcal{L}_{\protect\THETAVEC{}})}=19.9$. 
Likewise for the chain layout: for one set of minima (Fig. \ref{fig:D4_chain_NEB_50pts} top panel) the AUC was reduced from $2.825 \to 0.513$ and the maximum loss value for each NEB curve was reduced from $0.748 \to 0.123$ with $\max{(\max{\mathcal{L}_{\pi^{\prime}}}/\mathcal{L}_{\protect\THETAVEC{}})}=15.1$.  For the second set of minima (Fig. \ref{fig:D4_chain_NEB_50pts} (bottom panel)), the AUC was reduced from $1.542\to 0.326 $ and the maximum loss value for each NEB curve was reduced from $0.589 \to 0.078$ with $\max{(\max{\mathcal{L}_{\pi^{\prime}}}/\mathcal{L}_{\protect\THETAVEC{}})}=7.6$. 

For all landscapes, those centered around ravines or not, we see that the NEB can reduce the loss along piece-wise linear paths connecting minima.  Quantifying that change in loss can be done by comparing the reduction in maximum loss along a NEB path, or reduction in AUC for the NEB path.  This demonstrates how traversal of large barriers can be avoided using alternative paths (see the $D=1$ landscapes in Figs. \ref{fig:4P_chain_landscape},\ref{fig:4P_cycle_landscape}). We can also compare the maximum loss along a NEB path to the loss at the endpoints using $\max{\max{\mathcal{L}_{\pi^{\prime}}/\mathcal{L}_{\THETAVEC{}}}}$.  This value is difficult to minimize and shows that not all minima can be easily connected. The difficulty in finding a low loss path can be related to an inability to navigate barriers, for example the $D=2$ landscape in Fig. \ref{fig:7P_cycle_landscape}. Low loss paths can also be difficult to find due to the structure of the landscape, and the circuit design.  For example the $D=4$ landscape in Fig. \ref{fig:13P_cycle_landscape}, there are a number of irregular barriers that are encountered with small variations of the parameter vector-- it is possible that the gradient based updates in NEB could become trapped on local optima. Additionally, our PQC class used only a single rotation gate, $R_Y$.  This reduces the number of parameters to train, but with a trade-off that there are fewer degrees of freedom to explore the landscape.  Finally, we can also see that for some landscapes, the NEB path becomes flatter, and the maximum loss is reduced, but there appear to be wide saddle points between the minina, for example in the $D=4$ landscape in Fig. \ref{fig:D4_chain_NEB_50pts}. Yet we close by stating that most of the paths that we have found in our analysis have loss values that fall within the range of values plotted in Fig. \ref{fig:violin_plot}.

\section{Gate dropout stability}
\label{sec:dropout}
In classical machine learning, dropout stability is used to infer connectivity properties of the loss landscapes~\cite{shevchenko2020landscape}.  Here, we do not rigorously connect dropout stability to landscape connectivity, but we find that the orientation of ravines can be used to identify potential dropout patterns.    

While we draw on a number of methods developed in the classical machine learning literature, we do not draw any parallels between classical activation functions (neurons) and specific circuit elements.  Thus, we modify the definition of dropout.  In \cite{shevchenko2020landscape}, dropout stability is a quality of a trained neural network; a fraction of neurons in a trained network can be removed (by setting their outputs to 0) without significantly affecting the prediction quality (after a suitable rescaling of the remaining weights).  For PQCs, we redefine dropout as the single qubit gates in a trained PQC that can be replaced by identity gates ($\theta = 0$) without significantly affecting the prediction quality.  We refer to this as \textit{gate dropout}.

The ravines shown in Figs. \ref{fig:4P_chain_landscape} - \ref{fig:10P_cycle_landscape} are visualized by using planes defined by $3$ known minima (2 of which sit in a ravine). We chose to orient the landscapes such that any ravine will be nearly parallel with the $w_1$ axis. For $D=1$ and $D=2$ landscapes, translation along the $w_1$ direction can be done by updating a single parameter $\theta_i$; the support of vector $w_1$ is $\mathrm{supp}(w_1) = {\theta_i}$ meaning that once the training has converged to the ravine, updating $\theta_i$ parameter has minimal effect on the loss value and a possible gate dropout pattern would set $\theta_i = 0$. For the $D=1$ ravines shown in Figs. \ref{fig:4P_chain_landscape}-\ref{fig:4P_cycle_landscape}, $\theta_i = \theta_{2}$.  For the $D=2$ ravines shown in Fig. \ref{fig:7P_cycle_landscape}, $\theta_i = \theta_1$.
\begin{figure}[htbp]
  \centering
  \includegraphics[width=\columnwidth]{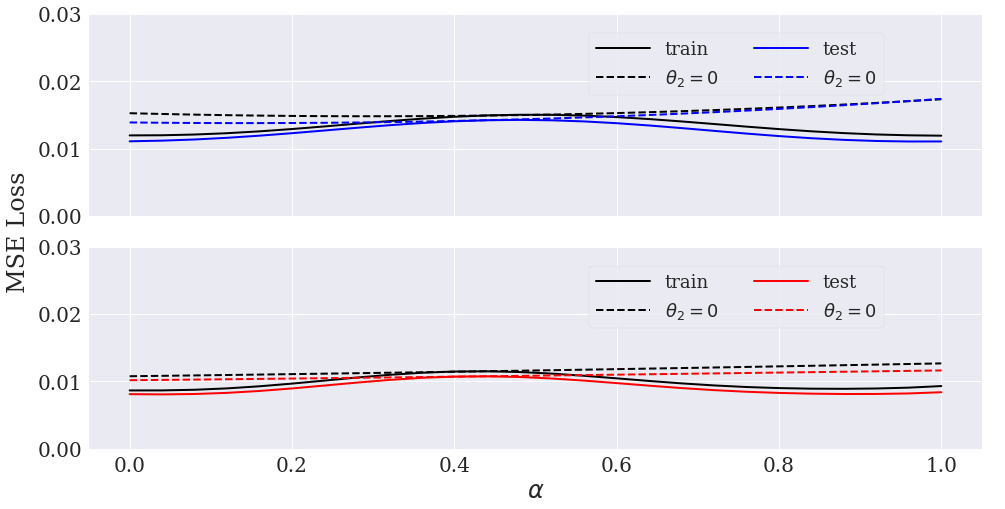}
  \caption{Change in loss value for a set of interpolated values in the $D=1$ landscapes for the chain layout (top panel) and the cycle layout (bottom panel). The solid lines show the loss values when all $4$ parameters are changed, the dash lines show the loss values when $\theta_2 = 0$ is fixed.}
  \label{fig:gate_dropout_D1}
\end{figure} 

\begin{figure}[htbp]
  \centering
  \includegraphics[width=\columnwidth]{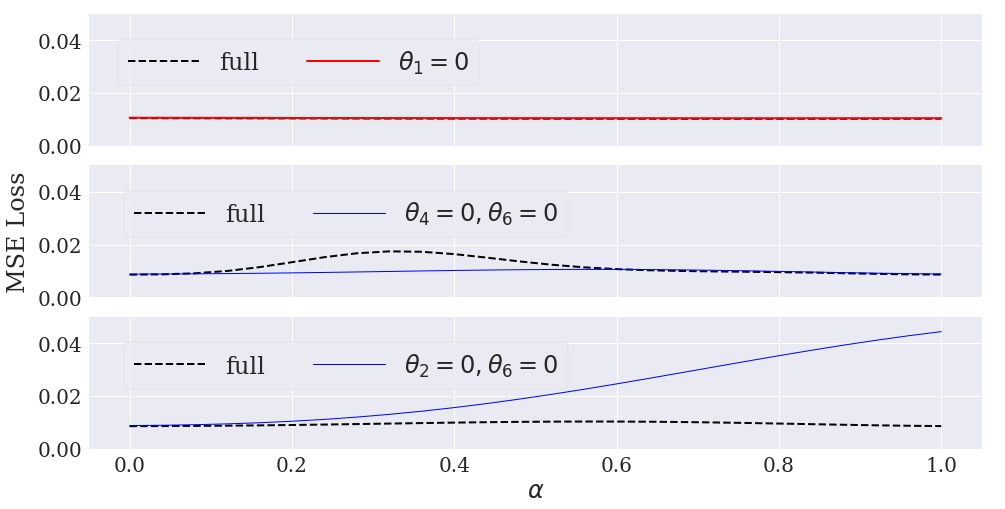}
  \caption{Change in loss value (evaluated on the training set only) for a set of interpolated values in the $D=2$ landscape for the cycle layout (top panel). The solid lines show the loss values when all $7$ parameters are changed, the dash lines show the loss values when $\theta_1 = 0$ is fixed.}
  \label{fig:dropout_tests}
\end{figure}

To test the gate dropout stability, we use the approach used in~\cite{shevchenko2020landscape}: for a given set of trained PQC parameters \THETAVEC{} we replace certain parameters with $0.0$, then re-evaluate the loss.  If the PQC is gate dropout stable, then the change in the loss should be minimal.  We can verify our gate dropout stability for the $D=1$ and $D=2$ circuits by showing the minima change in $\mathcal{L}$ evaluated along one-dimensional interpolated paths (see Figs. \ref{fig:gate_dropout_D1} and \ref{fig:dropout_tests} (top panel)).

However outside the trivial cases where translation along $w_1$ corresponds to a single parameter, determining a dropout pattern becomes difficult.  For $D=3$, $w_1$ depends on multiple directions. We test different dropout patterns by setting $\theta_2, \theta_6$ to zero, or $\theta_4,\theta_6$ to zero. In Fig. \ref{fig:dropout_tests} we see that fixing $\theta_2=0, \theta_6=0$ improves the loss along the straight line path.  However setting $\theta_4=0, \theta_6=0$ leads to a significant increase in the loss.

\section{Discussion}
\label{sec:dicussion}

During the course of this study we have executed nearly 100 training workflows for each combination of structural parameters:  entangling layer layout, optimizer, number of rotational gates, choice of initialization and batch size. Our results show wide variation in the training of these PQCs and this discussion will focus on the quality of the final trained PQC (quantified by the MSE on the test set). We first discuss several bounds on the MSE to frame our discussion of ``good'' and ``poor'' training, and highlight the limits of our chosen circuit ansatzae. We define the lower bound for the MSE using the noiseless curve.  A PQC which predicts values that lie exactly along the noiseless curve would result in a test set MSE of $0.003$.  Next, we set the upper bound for the MSE using examples of PQCs which fail to learn and return the same constant prediction irrespective of the input feature.  As the testing and training data is bounded by the range $[-\epsilon,1.0]$, then a PQC which predicts the value of $-1.0$ for all inputs will have the largest MSE on the test set of $1.78$.  In this study, no circuit reached the lower MSE bound-- the lowest MSE on the test set was $0.007$ which was observed for a circuit with $13$ parameters, chain entangling layers, batch size $8$, and trained with Adam.  The initial values were drawn from a Gaussian with $\mu = \frac{\pi}{2}$.  In this study, no circuit saturated the upper MSE bound-- the largest MSE on the test set was $0.156$ which was observed for a circuit with $7$ parameters, cycle entangling layers, batch size $2$, and trained with SGD. The initial values were drawn from a Gaussian with $\mu=0$.  We reiterate that the training of all circuits was done with a fixed number of optimization steps ($300$) and without early stopping.  

It is likely that for many PQCs, further optimization would improve the circuit training. However, fully saturating the bound on the MSE value is hindered by poor fitting of the y-values close to $+1$.  Improving the fit near these values can be achieved by incorporating post-processing of the measured value $\langle Z \rangle$.  In \cite{mitarai2018quantum} an additional constant was trained that re-scaled the output value $c \langle Z \rangle$.  Alternatively the training data could be re-scaled such that the predicted labels are constrained to a narrower range with zero mean $\lbrace y : |y|\ll 1\rbrace$. 

Our choice of circuit width and design was motivated by Ref. \cite{mitarai2018quantum}: it was shown that to fit a polynomial of degree $d$, then a circuit must have at least $d$ qubits. We initially assumed that a 3-qubit circuit has sufficient capacity to fit our target function $f(x) = x^2$.  On the other hand, Fig.~\ref{fig:quality_of_fit} shows that circuits with low test set MSE have difficulty fitting values close to $1.0$ or $0.0$. A larger number of qubits (6) with deeper circuits ($D = 6$) were shown to fit parabola $f(x) = x^2$ when the data was constrained to the range $[0.0, 1.0]$~\cite{mitarai2018quantum}.  Our results highlight the need for measures quantifying the expressibility and trainability of PQCs for different learning tasks.  These concepts have been studied with respect to general circuit expressibility \cite{sim2019expressibility}, quantum classifiers \cite{abbas2020power}, and general variational quantum algorithms \cite{holmes2021connecting}.   

In discussing mode connectivity, we agree with the following statements in \cite{kuditipudi2019connectivity}: the demonstration of connected low loss regions is not a guarantee that gradient based training will find such region. With $D=2$ circuits there is instability in the training performance: there are large outlier values for the MSE error and bi-modality in the MSE error distributions for $D=2$. Yet we know that there exist ravines in this landscape (see Fig. \ref{fig:7P_cycle_landscape}).

Though the training circuits are shallow, there are several noticeable differences in the minimum MSE values found with different entangling layer designs and circuits of different depth (Fig. \ref{fig:violin_plot}). With the chain layout the distribution of lowest MSE errors remains unimodal as the number of trainable parameters increases. Additionally, as the number of parameters is increased from $4$ to $7$, the median value of the lowest MSE error (for both layouts) nearly doubles, meaning that the likelihood of finding the region with lowest MSE error reduces with increasing parameter number. 

We close with a qualitative comparison of our $3$ optimizers.  From the values in \ref{tab:bin_values} we observe that QNG and Adam are far more adept at training PQCs to low test loss, compared to SGD. Fig. \ref{fig:violin_plot} and Table \ref{tab:bin_values} show that QNG and Adam frequently converge to regions of low MSE loss for both entangling layer designs. In contrast, training with SGD is able to return trained circuits with low MSE on the test set, but it also has a significant amount of outliers, leading to the conclusion that training is unstable in general with these layouts. In Fig. \ref{fig:violin_plot} it is clear that training with any of our chosen optimizers can have outliers (exhibited by long tails in the violin plot distributions).

\section{Conclusion}
\label{sec:conc}
The current discussion about trainability in variational quantum algorithms has largely centered on the occurrence of barren plateaus, and identifying characteristics of the loss landscape which will make training difficult.  Additionally, many studies have focused on landscapes with one global minimum (e.g. finding a ground state of a Hamiltonian).  Mode connectivity in PQCs is an interesting development, not only are there multiple minima in the loss landscape, but that connection between PQC circuit design and the geometry of the loss landscape goes beyond the periodicity of the single qubit gates.  In classical neural networks, the increase in loss along a connected path can be bounded for certain networks \cite{freeman2017topology}, but an analytic description of connected-ness in PQCs remains an open question. We quantify the connected-ness along a NEB path based on:  significant reduction in the maximum value of the loss along a NEB path, or reduction in the AUC for the NEB path.  Yet for many that value could still be a factor of 10 larger than the endpoints.  

Our study trained regression models using PQCs built with varying depth and  different entangling  layouts. Each model was trained via three  gradient-based  optimizers, and we compared  the  performance  by  the test set MSE and by examining the loss landscape for various circuit depths.  The  SGD  algorithm  in  general  was  less  stable  than  both  Adam and QNG, in terms of the variance in MSE error distributions. On the other hand, all optimizers showed the potential for trainability. Ravines in low dimensions do not serve as an impediment to training over higher dimensions in general (they may in fact be correlated with trainability). Regions of low loss are often interconnected with low-loss paths, as uncovered with NEB analysis, leading to the quantum analog of similar effects in classical deep learning. 

We have observed that mode connectivity can exist when the circuits have different entangling layer designs, but uncovering mode connectivity is dependent on how effectively the gradient-based NEB can explore the landscape. As with most brute force approaches, this experimental design will scale poorly to larger circuits (either increasing the number of parameters or number of qubits).  Increasing the number of parameters will increase the number of minima in the landscape; an exhaustive search of the landscape may require a prohibitively large number of independent training trials.  As the number of qubits increases, the training task may require a larger number of gradient steps in order to converge. 

Thus it remains to be seen if ravines and mode inter-connectivity persist in PQCs with larger widths and depths, which is a prime subject for future study. Our study was designed to demonstrate that the loss landscape associated with quantum circuit learning (supervised learning) is more complex then finding global minima. The visualization methods and modified NEB method used in this paper can be implemented for arbitrary circuit design, and should illuminate connectivity in larger dimensional systems and inform the trainability of a given PQC and optimizer combination. Ultimately, these tools may help the design of PQCs for QML tasks on NISQ hardware.

\section{Acknowledgements}
This work was partially supported as part of the ASCR Testbed Pathfinder Program at Oak Ridge National Laboratory under FWP ERKJ332.  This work was partially supported as part of the ASCR Fundamental Algorithmic Research for Quantum Computing Program at Oak Ridge National Laboratory under FWP ERKJ354. This work was partially supported as part of the ASCR QCAT Program at Oak Ridge National Laboratory under FWP \#ERKJ347. EL was supported by the U.S. Department of Energy, Office of Science, Office of Workforce Development for Teachers and Scientists (WDTS) under the Science Undergraduate Laboratory Internship program.

\section*{Conflict of interest}
 The authors declare that they have no conflict of interest.

\appendix
\section{Defining the AMS}
\label{appendix:clustering}
In the main text we noted that from multiple training runs we have multiple parameter sets that have low MSE loss on the test set.  These parameter sets are not identical and it is possible that multiple training runs will converge to the same minima in the landscape.  Yet we also know that the loss landscape of an over-parameterized model will contain multiple minima.  To identify the regions that may correspond to different minima in the loss landscape, we apply mean shift clustering \cite{comaniciu2002mean} implemented in the Python library \texttt{scikit-learn} \cite{scikit-learn}.  We apply the clustering on a subset of parameter vectors:  the trainings which have converged to the lowest MSE values (see Table \ref{tab:bin_values} in the main text). From these values we define a subset of minima called the \textit{aggregate minima set}.
\begin{table}[h!]
  \centering
 \caption{Number of parameter sets in lowest MSE bin ($n_p$), unique cluster centers found by the mean shift algorithm ($n_c$) and the cardinality of the AMS ($n^{*}_c$).}
\begin{tabular}{| c | c | c | c | c |}
 \hline
Layout & Gates & $n_{p}$  & $n_{c}$  & $|\mathrm{AMS}|$\\
 \hline
 \hline
 \multirow{4}*{Cycle}
&4  &  244 &  16 &  14 \\
&7  &  67 &  19 &  18 \\
&10 &  164 &  28 &  24 \\
&13 &  120 &   13 &  10 \\
\cline{2-5}
\hline
\hline
 \multirow{4}*{Chain}
&4  &  279 &  20 &  20\\
&7  &  155 &  25 &  7\\
&10 &  194 &  20 &  19\\
&13 &  192 & 14 &  12\\
\cline{2-5}
\hline
\end{tabular}
\label{tab:cluster_values}
\end{table}

In the main text we apply the nudged elastic band analysis to the mean shift cluster centers which correspond to low loss regions.  By using mean shift clustering, we do not restrict the number of unique cluster centers that can be found in the data.  For each circuit design (depth and layout) mean shift clustering will find a different number of unique cluster centers, but it is also possible that the algorithm will assign a cluster center that does not sit in a low loss region (meaning that the algorithm attempts to assign a cluster center between two sparsely populated minima).  Thus after mean shift is applied to the parameter subsets, we find $n_c$ clusters, and then by evaluating the loss at each cluster center, we define the final AMS by discarding any center that does not return a MSE that sits in the lowest bin.

\section{Quantum circuit NEB}
\label{appendix:quantum_NEB}
The Nudged Elastic Band algorithm (NEB) \cite{henkelman2000neb} was first implemented as a search method for minimum energy paths in chemical potential landscapes.  With a known potential function $V(\mathbf{R})$ which is a function of an distance parameter $\mathbf{R}$, an initial proposed minimum energy path is defined by a set of $N$ pivot points $\lbrace R_i \rbrace$ (the endpoints are known minima and remain fixed).  Each pivot point is updated by a tangential and parallel forces acting on it (Eqtns  3 and 4 of Ref.~\cite{henkelman2000neb}),
\begin{equation}
\mathbf{F}_i = \mathbf{F}_i^{S}|_{\Vert} - \nabla V(\mathbf{R}_i)|_{\perp}.
\label{eq:neb_updates}
\end{equation}
These forces can be defined with respect to parameterized quantum circuit models by using the loss function $\mathcal{L}$ and the circuit parameters \THETAVEC.

A pivot point \THETAVEC{i} has neighboring pivot points \THETAVEC{i-1} and \THETAVEC{i+1}. The local tangents $\mathbf{\tau})i$ are evaluated in the parameter space using normalized line segments using the neighboring point with higher loss: $\mathbf{\tau}_i = \THETAVEC{i+1} - \THETAVEC{i}$ if $\mathcal{L}(\THETAVEC{i+1})>\mathcal{L}(\THETAVEC{i-1})$, or $\mathbf{\tau}_i = \THETAVEC{i+1} - \THETAVEC{i}$ if $\mathcal{L}(\THETAVEC{i-1})>\mathcal{L}(\THETAVEC{i+1})$.  The local tangents are used to compute the parallel term ($\mathbf{F}_i^{S}|_{\Vert}$), which ensures that the pivot points remain separated along the path, and the tangential force ($\nabla \mathcal{L}(\THETAVEC{i})|_{\perp}$) moves pivots to regions of lower energy. Equations 4 and 12 of Ref.~\cite{henkelman2000neb} can be recast in terms of \THETAVEC{} and $\mathcal{L}$.  Our version of NEB was written and implemented using the PennyLane library \cite{bergholm2020pennylane}.

\end{document}